\documentclass[10pt,letterpaper,twocolumn]{article} 

\usepackage{ol2}
\usepackage[draft]{hyperref}
\usepackage{amsmath}
\usepackage{epstopdf}

\begin{document}

\twocolumn[ 

\title{Experimental demonstration of terahertz metamaterial absorbers with a broad and flat high absorption band}


\author{Li Huang,$^{1,2,*}$ Dibakar Roy Chowdhury,$^2$ Suchitra Ramani,$^2$ Matthew T. Reiten,$^2$ Sheng-Nian Luo,$^3$ Antoinette J. Taylor,$^2$ and Hou-Tong Chen$^{2,*}$}

\address{
$^1$Physics Department, Harbin Institute of Technology, Harbin, Heilongjiang 150001, China\\
$^2$Los Alamos National Laboratory, MPA-CINT, MS K771, Los Alamos, New Mexico 87545, USA\\
$^3$Los Alamos National Laboratory, P-25, MS H846, Los Alamos, New Mexico 87545, USA\\
$^*$Corresponding authors: lihuang2002@hit.edu.cn; chenht@lanl.gov}

\begin{abstract}We present the design, numerical simulations and experimental measurements of THz metamaterial absorbers with a broad and flat absorption top both for transverse electric and transverse magnetic polarizations over a wide incidence angle range. The metamaterial absorber unit cell consists of two sets of structures resonating at different but close frequencies. The overall absorption spectrum is the superposition of individual components and becomes flat at the top over a significant bandwidth. The experimental results are in excellent agreement with numerical simulations.
\end{abstract}


 ] 

\noindent Since the first demonstration by Landy et al.~\cite{Landy2008PRL}, metamaterial absorbers have attracted a great deal of interest worldwide during the past few years due to a host of potential applications including detectors, imaging, and sensing. They are typically comprised two structured metallic layers separated with a dielectric spacer, either free standing or supported by a suitable substrate, with a total thickness of functional layers much smaller than the operational wavelength. The originally proposed bottom structured metallic layer was a resonant cut-wire array~\cite{Landy2008PRL,Tao2008OE}, which has evolved to a simpler metal ground plane~\cite{Tao2008PRB}. This makes the metamaterial absorbers more or less resemble Salisbury screens or circuit analog absorbers~\cite{Munk2000FSS}. A variety of metamaterial structures have been employed and the operational frequency has covered the microwave~\cite{Landy2008PRL} through terahertz (THz)\cite{Tao2008OE,Tao2008PRB,Shchegolkov2010PRB} to optical~\cite{Liu2010RPL,Liu2010NL} ranges. The generally accepted idea was that by tuning the effective electric permittivity $\epsilon$ and magnetic permeability $\mu$ independently, it is possible to realize impedance matching to free-space~\cite{Landy2008PRL} and minimize the reflection. However, it has been recently found that a Fabry-Per\'{o}t resonance between the two metallic layers is responsible for the observed metamaterial absorption~\cite{Chen2010PRL}, where the tuned reflection/transmission amplitude and phase by the metamaterial layers satisfy the antireflection requirements similar to a quarter wave antireflection, i.e. another kind of impedance matching. We have verified this mechanism for many metamaterial absorbers proposed in literature~\cite{Chen2011LPR}.

Besides its polarization and incidence angle dependence, the bandwidth of a metamaterial absorber is one of the important aspects that may affect many applications. So far, most designs of metamaterial absorbers operate at a specific narrow frequency range. Dual-band~\cite{Wen2009APL,Tao2010JPD,Ma2011OL} and triple-band~\cite{Shen2011OE ,Li2011JAP} metamaterial absorbers have been demonstrated with distinct narrow absorption frequencies; however, broadband metamaterial absorbers remain a challenge and there have been only a few very recent studies, mostly focusing on theoretical and numerical investigation~\cite{Hu2009OE,Ye2010JOSAB,Wakatsuchi2010OE,Luo2011EPJB,Alici2011OE, Grant2011OL}. In this Letter, we experimentally demonstrate metamaterial absorbers operating in the THz frequency range with a broad and flat high absorption top over a wide incidence angle range, which are in excellent agreement with numerical simulations.

\begin{figure}[t!]
\centerline{\includegraphics[width=3in]{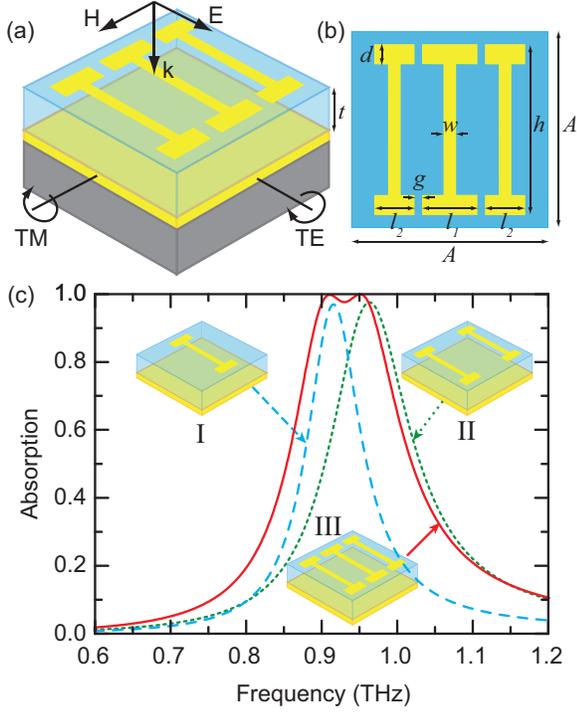}} \caption{Schematic of the whole unit cell (a) and top view (b) of the metamaterial absorber with dimensions (in $\mu$m) $A=78$, $h = 68$, $d = 8$, $w = 5$, $t = 8.5$, $g=2.5$, $l_1 = 21.5$, and $l_2 = 16.5$ or 17.5. TE and TM polarizations are also indicated in (a). (c) Numerical simulation results of absorption spectra at normal incidence for three different configurations of the I-shaped resonators which are indicated in the insets.}
\label{Fig1}
\end{figure}

The schematic of THz metamaterial absorber unit cell is shown in Figs.~\ref{Fig1}(a) and \ref{Fig1}(b), which comprises three I-shaped resonators separated from a ground plane using a dielectric spacer. It is symmetrically designed that the two side I-shaped resonators are identical and differ from the center resonator only in the length of the end loading, i.e. $l_1 = 21.5$~$\mu$m and $l_2 = 16.5$ or 17.5~$\mu$m in our two designs. Finite-element numerical simulations were first performed with CST Microwave Studio 2009 using periodic boundary conditions and under normal incidence, for the electric field parallel to the long axis of the resonators (Fig.~\ref{Fig1}(a)). The absorptivity $A(\omega)$ was then obtained from the S-parameters by $A(\omega)=1- R(\omega)- T(\omega)=1-|S_{11}|^2-|S_{12}|^2$, where $T(\omega)$ is very close to zero when using a gold ground plane of 200~nm thickness, which is the same as in the I-shaped resonators. Three differently configured metamaterial absorbers, as shown in the insets to Fig.~\ref{Fig1}(c), were simulated using a relative dielectric constant $\epsilon_s = 3.1$ and loss $\tan\delta = 0.07$ for the dielectric spacer (polyimide) of thickness $t = 8.5$~$\mu$m. All of them exhibit strong absorption of THz radiation with a peak absorptivity close to unity. The absorption peak is narrow for the configurations with either center (configuration I) or side (configuration II) I-shaped resonators alone. However, deploying the two types of resonators in one unit cell (configuration III) results in a broader absorption top, a superposition of configurations I and II. Further simulations show that the bandwidth increases as the frequency difference between the absorption peaks of I and II increases; however, the small dip in III also becomes undesirably wider and deeper. On the other hand, when the absorption peaks of I and II are closer, the absorption top is flatter but with a reduced overall bandwidth.

Two metamaterial absorbers of configuration III were fabricated using parameters shown in the caption of Fig.~\ref{Fig1}. First, the ground plane was deposited by e-beam evaporation of 10-nm-thick Ti, 200-nm-thick Au, and then again 10-nm-thick Ti on a GaAs substrate.  The dielectric spacer was formed using a spin-coated and thermally cured polyimide layer with a final thickness of $\sim$$8~\mu$m. Finally, the I-shaped resonator array was patterned using conventional photolithography methods, e-beam deposition of 10-nm-thick Ti and 200-nm-thick Au, and lift-off processes. Note that the purpose of the Ti layers is to increase the adhesion of the Au layers to the GaAs substrate and polyimide spacer. The metamaterial absorbers were characterized by reflection measurements using a fiber-coupled THz time-domain spectrometer in reflection mode~\cite{Chen2010PRL,OHara2007JNO} at various incidence angles between 30$^\circ$ and 60$^\circ$, using a blank Au coated substrate as the reference. The absorptivity was obtained by $A(\omega)=1- R(\omega)$ since $T(\omega)$ is zero.

\begin{figure}[h!]
\centerline{\includegraphics[width=3in]{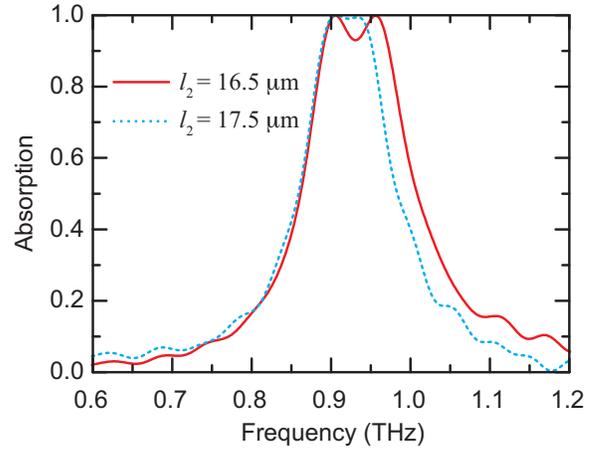}} \caption{Measured THz absorption spectra of two metamaterial absorbers with $l_2 = 16.5~\mu$m (solid red curve) and 17.5~$\mu$m (dotted blue curve) for TE polarization and at a 30$^\circ$ incidence angle.}
\label{Fig2}
\end{figure}

The absorption spectrum of the metamaterial absorber with $l_2 = 16.5~\mu$m (sample \#1) is shown as the solid curve in Fig.~\ref{Fig2}, in agreement with the prediction in Fig.~\ref{Fig1}. It was measured at an incidence angle of $30^\circ$ and with the THz electric field parallel to the resonators, i.e. transverse electric (TE) polarization. The highest absorptivity measured is 99.9\% at 0.905 and 0.956 THz, between which the smallest absorptivity in the dip is still as high as 93\%. The width of the absorption top is 0.1 THz even if we define a strict criterion of 80\% of the maximum absorption, and the roll-off is rather fast for broadband operation. As described before, we can largely eliminate the absorption dip by making a smaller frequency difference between the absorption peaks in configurations I and II. This is clearly verified by the dotted absorption spectrum in Fig.~\ref{Fig2} obtained from a second metamaterial absorber (sample \#2) by slightly increasing $l_2 = 17.5~\mu$m, which reduces the absorption peak frequency of configuration II, while all other parameters are kept the same as in sample \#1. Between 0.899 THz and 0.939~THz (a width of 40~GHz) the absorptivity remains higher than 99\%, although the width of the absorption top, as defined above, is slightly reduced to 0.08~THz.

\begin{figure}[t!]
\centerline{\includegraphics[width=3in]{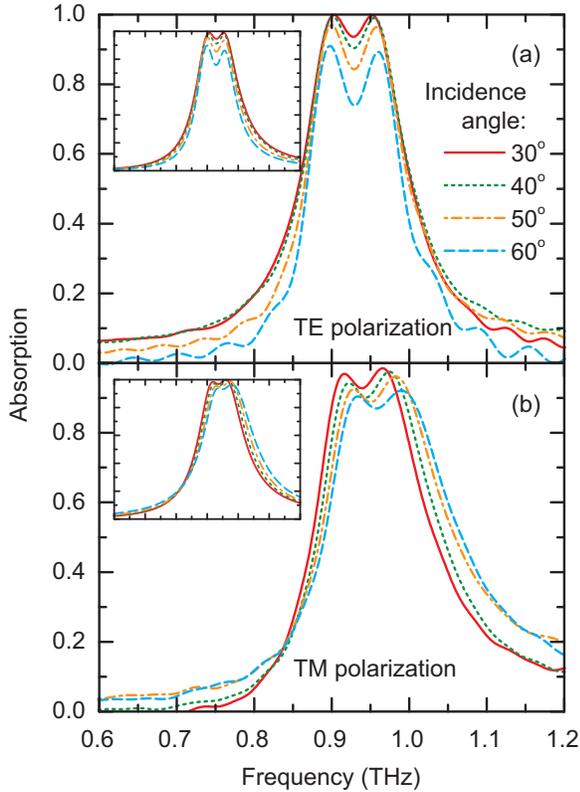}} \caption{Experimentally measured absorption spectra for (a) TE and (b) TM polarizations at various incidence angles from 30$^{\circ}$ to 60$^{\circ}$ with an increment of 10$^{\circ}$. Insets: simulated absorption spectra for the respective polarizations and incidence angles.}
\label{Fig3}
\end{figure}

We further characterized the incidence angle dependence of the absorption for the two metamaterial absorber samples, and the results for sample \#1 are shown in Fig.~\ref{Fig3} for both the transverse electric (TE) and transverse magnetic (TM) polarizations (indicated in Fig.~\ref{Fig1}(a)) of the incident THz radiation. The experimental results (not shown) for sample \#2 exhibit similar behavior. Increasing the incident angle results in an overall decrease of the absorptivity, and the dip becomes deeper particularly in the absorption spectra for TE polarization. At the large incident angle of $60^\circ$, the peak absorptivity is above 90\%, and the smallest absorptivity in the absorption dip is still as high as 74\% for TE and 87\% for TM polarization. The absorption band does not shift with the incidence angle for TE polarization, and it shifts only slightly for TM polarization. For both polarizations the width of the absorption top does not change significantly. The corresponding numerical simulation results are shown in the insets to Fig.~\ref{Fig3} for the respective polarizations and incidence angles, which are in excellent agreement with experimental measurements. Additional numerical simulations (results not shown) also reveal that at incidence angles smaller than $30^\circ$ the absorption spectra have a very small variation.

In summary, we have designed and experimentally demonstrated THz metamaterial absorbers with a broad and flat absorption top over a wide incidence angle range for both TE and TM polarized THz radiation. The experimental results are in excellent agreement with numerical simulations. Further expansion of the absorption bandwidth is possible by improving the metamaterial design, and the concept could be readily extended to other frequency regimes for a host of applications that have a wide bandwidth requirement.

L.H. acknowledges support from Natural Science Foundation of China (NSFC) under Grant No. 10904023. We acknowledge support from the Los Alamos National Laboratory LDRD program. This work was performed, in part, at the Center for Integrated Nanotechnologies, a US Department of Energy, Office of Basic Energy Sciences Nanoscale Science Research Center operated jointly by Los Alamos and Sandia National Laboratories. Los Alamos National Laboratory, an affirmative action/equal opportunity employer, is operated by Los Alamos National Security, LLC, for the National Nuclear Security Administration of the US Department of Energy under contract DE-AC52-06NA25396.

\bibliographystyle{osajnl}

\newpage

\bibliographystyle{osajnl}

\begin{thebibliography}{99}
\bibitem{Landy2008PRL} N. I. Landy, S. Sajuyigbe, J. J. Mock, D. R. Smith, and W. J. Padilla, ``Perfect metamaterial absorber'', Phys. Rev. Lett. \textbf{100}, 207402 (2008).
\bibitem{Tao2008OE} H. Tao, N. I. Landy, C. M. Bingham, X. Zhang, R. D. Averitt, and W. J. Padilla, ``A metamaterial absorber for the terahertz regime: Design, fabrication and characterization'', Opt. Express \textbf{16}, 7181 (2008).
\bibitem{Tao2008PRB} H. Tao, C. M. Bingham, A. C. Strikwerda, D. Pilon, D. Shrekenhamer, N. I. Landy, K. Fan, X. Zhang, W. J. Padilla, and R. D. Averitt, ``Highly flexible wide angle of incidence terahertz metamaterial absorber: Design, fabrication, and characterization'', Phys. Rev. B \textbf{78}, 241103 (2008).
\bibitem{Munk2000FSS} B. A. Munk, \textit{Frequency Selective Surfaces: Theory and Design} (Wiley, 2000).
\bibitem{Shchegolkov2010PRB} D. Y. Shchegolkov, A. K. Azad, J. F. O'Hara, and E. I. Simakov, ``Perfect subwavelength fishnetlike metamaterial-based film terahertz absorbers'', Phys. Rev. B \textbf{82}, 205117 (2010).
\bibitem{Liu2010RPL} X. Liu, T. Starr, A. F. Starr, and W. J. Padilla, ``Infrared spatial and frequency selective metamaterial with near-unity absorbance'', Phys. Rev. Lett. \textbf{104}, 207403 (2010).
\bibitem{Liu2010NL} N. Liu, M. Mesch, T. Weiss, M. Hentschel, and H. Giessen, ``Infrared perfect absorber and its application as plasmonic sensor'', Nano Lett. \textbf{10}, 2342 (2010).
\bibitem{Chen2010PRL} H.-T. Chen, J. F. Zhou, J. F. O'Hara, F. Chen, A. K. Azad, and A. J. Taylor, ``Antireflection coating using metamaterials and identification of its mechanism'', Phys. Rev. Lett. \textbf{105}, 073901 (2010).
\bibitem{Chen2011LPR} H.-T. Chen, J. F. O'Hara, A. K. Azad, and A. J. Taylor, ``Manipulation of terahertz radiation using metamaterials'', Laser Photon. Rev. \textbf{5}, 513 (2011).
\bibitem{Wen2009APL} Q.-Y. Wen, H.-W. Zhang, Y.-S. Xie, Q.-H. Yang, and Y.-L. Liu, ``Dual band terahertz metamaterial absorber: Design, fabrication, and characterization'', Appl. Phys. Lett. \textbf{95}, 241111 (2009).
\bibitem{Tao2010JPD} H. Tao, C. M. Bingham, D. Pilon, K. Fan, A. C. Strikwerda, D. Shrekenhamer, W. J. Padilla, X. Zhang, and R. D. Averitt, ``A dual band terahertz metamaterial absorber'', J. Phys. D: Appl. Phys. \textbf{43}, 225102 (2010).
\bibitem{Ma2011OL} Y. Ma, Q. Chen, J. Grant, S. C. Saha, A. Khalid, and D. R. S .Cumming, ``A terahertz polarization insensitive dual band metamaterial absorber'', Opt. Lett. \textbf{36}, 945 (2011).
\bibitem{Li2011JAP} H. Li, L. H. Yuan, B. Zhou, X. P. Shen, Q. Cheng, and T. J. Cui, ``Ultrathin multiband gigahertz metamaterial absorbers'', J. Appl. Phys. \textbf{110}, 014909 (2011).
\bibitem{Shen2011OE} X. Shen, T. J. Cui, J. Zhao, H. F. Ma, W. X. Jiang, and H. Li, ``Polarization-independent wide-angle triple-band metamaterial absorber'', Opt. Express \textbf{19}, 9401 (2011).
\bibitem{Hu2009OE} C. Hu, L. Liu, Z. Zhao, X. Chen, and X. Luo, ``Mixed plasmons coupling for expanding the bandwidth of near-perfect absorption at visible frequencies'', Opt. Express \textbf{17}, 16745 (2009).
\bibitem{Ye2010JOSAB} Y. Q. Ye, Y. Jin, and S. He, ``Omnidirectional, polarization-insensitive and broadband thin absorber in the terahertz regime'', J. Opt. Soc. Am. B \textbf{27}, 498 (2010).
\bibitem{Wakatsuchi2010OE} H. Wakatsuchi, S. Greedy, C. Christopoulos, and J. Paul, ``Customised broadband metamaterial absorbers for arbitrary polarisation'', Opt. Express \textbf{18}, 22187 (2010).
\bibitem{Luo2011EPJB} H. Luo, Y. Z. Cheng, and R. Z. Gong, ``Numerical study of metamaterial absorber and extending absorbance bandwidth based on multi-square patches'', Eur. Phys. J. B \textbf{81}, 387 (2011).
\bibitem{Alici2011OE} K. B. Alici, A. B. Turhan, C. M. Soukoulis, and E. Ozbay, ``Optically thin composite resonant absorber at the near-infrared band: a polarization independent and spectrally broadband configuration'', Opt. Express \textbf{19}, 14260 (2011).
\bibitem{Grant2011OL} J. Grant, Y. Ma, S. Saha, A. Khalid, and D. R. S. Cumming, ``Polarization insensitive, broadband terahertz metamaterial absorber'', Opt. Lett. \textbf{36}, 3476 (2011).
\bibitem{OHara2007JNO} J. F. O'Hara, E. Smirnova, H.-T. Chen, A. J. Taylor, R. D. Averitt, C. Highstrete, M. Lee, and W. J. Padilla, ``Properties of planar electric matamaterials for novel terahertz applications'', J. Nanoelectron. Optoelectron. \textbf{2}, 90 (2007).

\end{thebibliography}

\end{document}